\documentclass[12pt]{iopart}

\usepackage[dvips]{graphicx}
\usepackage{amsbsy}
\begin{document}

\title[Structural and spectral properties of a family of deterministic recursive trees]{Structural and spectral properties of a family of deterministic recursive trees: Rigorous solutions}

\author{Yi Qi$^{1,2}$, Zhongzhi Zhang$^{1,2}$, Bailu Ding$^{1,2}$,  Shuigeng Zhou$^{1,2}$,  and Jihong Guan$^{3}$}
\address{$^1$School of Computer Science, Fudan
University, Shanghai 200433, China\\
$^2$Shanghai Key Lab of Intelligent Information Processing, Fudan
University, Shanghai 200433, China\\
$^3$Department of Computer Science and Technology, Tongji
University, 4800 Cao'an Road, Shanghai 201804, China}

\ead{zhangzz@fudan.edu.cn,sgzhou@fudan.edu.cn,jhguan@tongji.edu.cn}

\begin{abstract}
As one of the most significant models, the uniform recursive tree
(URT) has found many applications in a variety of fields. In this
paper, we study rigorously the structural features and spectral
properties of the adjacency matrix for a family of deterministic
uniform recursive trees (DURTs) that are deterministic versions of
URT. Firstly, from the perspective of complex networks, we
investigate analytically the main structural characteristics of
DURTs, and obtain the accurate solutions for these properties, which
include degree distribution, average path length, distribution of
node betweenness, and degree correlations. Then we determine the
complete eigenvalues and their corresponding eigenvectors of the
adjacency matrix for DURTs. Our research may shed light in better
understanding of the features for URT. Also, the analytical methods
used here is capable of extending to many other deterministic
networks, making the precise computation of their properties
(especially the full spectrum characteristics) possible.
\end{abstract}

%Uncomment for PACS numbers title message
% Keywords required only for MST, PB, PMB, PM, JOA, JOB?
%  Small-world networks\sep Disordered systems\sep  Networks
%\vspace{2pc}
%\noindent{\it Keywords}: Article preparation, IOP journals
% Uncomment for Submitted to journal title message
%\submitto{\JPA}
% Comment out if separate title page not required

%89.20.Hh World Wide Web, Internet
%89.75.Da Systems obeying scaling laws
%89.75.Fb Structures and organization in complex systems
%89.75.-k Complex systems
%89.75.Hc Networks and genealogical trees

\maketitle

%%%%%%%%%%%%%%%%%%%%%%%%%%%%%%%%%%%%%%%%%%%%%%%%%%%%%%%%%%%%%%%%%
%%%%%%%%%%%%%%%%%%%%%%%%%%%%%%%%%%%%%%%%%%%%%%%%%%%%%%%%%%%%%%%%%
%\vskip -0.5cm\color{Blue}
%\vbox to 0pt{\kern -14cm {
%\noindent \small \copyright 2005
%{\em  All rights reserved}\\
%{\em J of Physica A}, submitted.}
%\vss}\color{Black}

\section{Introduction}

Structural characterization is very significant for the study in the
field of complex networks that have become a focus of attention for
the scientific community~\cite{AlBa02,DoMe02,Ne03}. In the past
decade, great efforts have been dedicated to characterizing and
understanding the structural properties of real
networks~\cite{CoRoTrVi07}, including degree distribution, average
path length (APL), betweenness, degree correlations, fractality, and
so forth. These measures have a profound effect on various dynamical
processes taking place on top of complex networks~\cite{DoGoMe08},
such as robustness~\cite{AlJeBa00,CaNeStWa00,CoErAvHa01,ZhZhZo07},
epidemic spreading~\cite{PaVe01a,ZhZhZoCh08},
synchronization~\cite{BaPe02,ZhRoZh06,CoGa07,ArDiKuMoZh08}, and
games~\cite{SzFa07}.

The foregoing measurements focus on direct measurements of
structural properties of networks, and play an important role in
understanding network complexity~\cite{Ba05}. Aside from these
measurements there exists a vast literature related to spectrum of
complex networks~\cite{FaDeBaVi01,GoKaKi01,DoGoMeSa03,ChLuVu03},
which provides useful insight into the relevant structural
properties of and dynamical processes on graphs. In contrast to the
fact that structural features capture the static topological
properties of complex networks, spectra (eigenvalues and
eigenvectors) of adjacency matrix provide global measures of the
characterization for network topology. In a variety of dynamical
processes, the impact of network structure is encoded in the spectra
of its adjacency matrix, especially the extreme eigenvalues and
their corresponding eigenvectors. For example, in the dynamical
model for the spreading of infections, the epidemic thresholds are
governed by the largest eigenvalue of the adjacency
matrix~\cite{BoPaVe03,ChWaWaLeFa08}, which also plays a fundamental
role in determining critical couplings for the onset of coherent
behavior~\cite{ReOtHu06}. In addition, recent research showed that
in the Susceptible-Infected model of epidemic outbreaks on complex
networks, the eigenvector corresponding to the largest eigenvalue
are related to spreading power of network nodes~\cite{CaEnCo06}. In
spite of the importance of the eigenvalues and eigenvectors of the
adjacency matrix, however, until now, most analysis of the spectra
has been confined to approximate or numerical methods, the latter of
which is prohibitively time and memory consuming for large-scale
networks~\cite{FaDeBaVi01}.

On the other hand, in order to mimic real systems and study their
structural properties, a great number of network models have been
presented~\cite{AlBa02,DoMe02,Ne03}, among which the uniform
recursive tree (URT) is perhaps one of the most widely studied
models~\cite{SmMa95}. It is now established that the URT, together
with the famous Erd\"os-R\'enyi model~\cite{ErRe60}, constitutes the
two principal models~\cite{DoKrMeSa08,ZhZhZhGu08} of random graphs.
As one of the simplest trees, the URT is constructed as follows:
start with a single node, at each time step, we attach a new node to
an existing node selected at random. It has found many important
applications in various areas. For example, it has been suggested as
models for the spread of epidemics~\cite{Mo74}, the family trees of
preserved copies of ancient or medieval texts~\cite{NaHe82}, chain
letter and pyramid schemes~\cite{Ga77}, to name but a few.

Recently, a class of deterministically growing tree-like networks
have been proposed to describe real-world systems whose number of
nodes increases exponentially with time~\cite{JuKiKa02}. We call
them deterministic uniform recursive trees (DURTs), since they are
deterministic versions of URT. This kind of deterministic models
have received considerable attention from the scientific communities
and have turned out to be a useful
tool~\cite{BaRaVi01,DoGoMe02,CoFeRa04,ZhRoZh07,RaBa03,NaUeKaAk05,AnHeAnSi05,ZhRoCo06,CoOzPe00,ZhRoGo06,Hi07,BaCoDa06,ZhZhFaGuZh07,ZhZhZoChGu07,BeMaRo08,BoGoGu08,BaCoDaFi08,ZhZhQiGu08}.
Although uniform recursive tree is well
understood~\cite{SmMa95,DoKrMeSa08,ZhZhZhGu08,Mo74,NaHe82,GoOhJeKaKi02,DoMeOl06},
relatively less is known about the structural and other nature of
the DURTs~\cite{JuKiKa02}.

In this paper, from the viewpoint of complex networks, we offer a
comprehensive analysis of the deterministic uniform recursive trees
(DURTs)~\cite{JuKiKa02}. We firstly determine exactly relevant
structural properties of the DURTs, including degree distribution,
average path length, betweenness distribution, and degree
correlations. Then, using methods of graph theory and algebra, We
calculate all the eigenvalues and eigenvectors of the adjacency
matrix, which are obtained through the recurrence relations derived
from the very structure of the DURTs.

\section{The deterministic uniform recursive trees}

\begin{figure}
\begin{center}
\includegraphics[width=0.6\linewidth,trim=60 180 60 340]{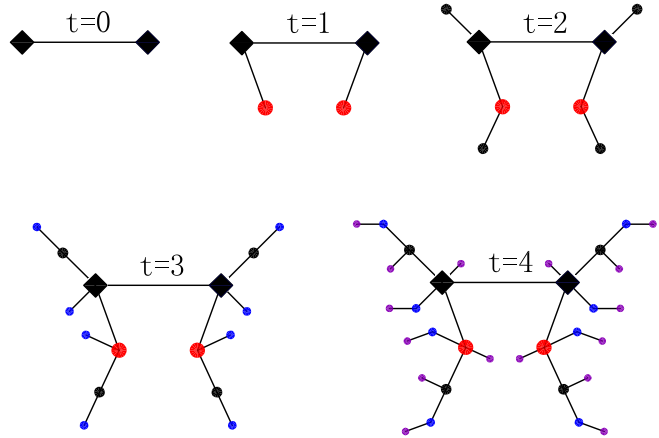}
\caption{Illustration of a deterministic uniform recursive tree for
the special case of $m=1$, showing the first several steps of growth
process.} \label{net01}
\end{center}
\end{figure}

The deterministic uniform recursive trees under consideration are
constructed in an iterative way~\cite{JuKiKa02}. We denote the trees
(networks) after $t$ steps by $U_{t}$ ($t\geq 0$). Then the networks
are built as follows. For $t=0$, $U_{0}$ is an edge connecting two
nodes. For $t\geq 1$, $U_{t}$ is obtained from $U_{t-1}$. We attach
$m$ new nodes to each node in $U_{t-1}$. This iterative process is
repeated, then we obtain a class of deterministically growing trees
with an exponential decreasing spectrum of degrees as shown below.
The definition of the model for a particular case of $m=1$ is
illustrated schematically in figure~\ref{net01}.

%%%%%%%%%%%%%%%%%%%%%%%%%%%%%%%%%%%%%%%%%%%%%%%%%%%%%%%%%%
% Figure  1
%%%%%%%%%%%%%%%%%%%%%%%%%%%%%%%%%%%%%%%%%%%%%%%%%%%%%%%%%%
%%%%%%%%%%%%%%%%%%%%%%%%%%%%%%%%%%%%%%%%%%%%%%%%%%%%%%%%%%

We first compute the total number of nodes $N_t$ and the total
number of edges $E_t$ in the $U_{t}$. Let $n_v(t)$ and $n_e(t)$
denote the numbers of nodes and edges created at step $t$,
respectively. Then, $n_v(0)=N_{0}=2$ and $n_e(0)=1$. By
construction, we have $n_v(t)=m\,N_{t-1}$, thus
$N_{t}=n_v(t)+N_{t-1}=(1+m)\,N_{t-1}$. Considering the initial
condition $N_{0}=2$, we obtain $N_t=2\,(1+m)^{t}$ and
$n_v(t)=2\,m\,(1+m)^{t-1}$. Thus, $E_t=N_t-1=2\,(1+m)^{t}-1$. Notice
that at arbitrary step $t\geq 1$, the addition of each new node
leads to only new edge, so $n_e(t)=n_v(t)=2\,m\,(1+m)^{t-1}$ for all
$t\geq 1$.

\section{Structural properties}

In this section, we investigate four important structural properties
of $U_{t}$, including degree distribution, average path length,
betweenness distribution, and degree correlations.

\subsection{Degree distribution}

By definition, the degree of a node $i$ is the number of edges
connected to $i$. The degree distribution $P(k)$ of a network is the
probability that a randomly selected node has exactly $k$ edges. Let
$k_{i}(t)$ denote the degree of node $i$ at step $t$. If node $i$ is
added to the network at step $t_i$, then by construction, $k_{i}(
t_i)=1$. In each of the subsequent time steps, $m$ new nodes will be
created connected to $i$. Thus the degree $k_i(t)$ of node $i$
satisfies the relation
\begin{equation}
k_{i}(t)=k_{i}(t-1)+m.
\end{equation}
Considering the initial condition $k_{i}( t_i)=1$, we obtain
\begin{equation}\label{Ki}
k_{i}(t)=1+m\,(t-t_{i}).
\end{equation}
Since the degree of each node has been obtained explicitly as in
equation~(\ref{Ki}), we can get the degree distribution via its
cumulative distribution~\cite{Ne03}
\begin{equation} \label{cumulative distribution1}
P_{cum}(k)=\sum_{k'=k}^{\infty}P(k'),
\end{equation}
which is the probability that the degree is greater than or equal to
$k$. An important advantage of the cumulative distribution is that
it can reduce the noise in the tail of probability distribution.
Moreover, for some networks whose degree distributions have
exponential tails: $P(\tilde{k}) \sim e^{-\tilde{k}/\kappa}$, the
cumulative distribution also has an exponential expression with the
same exponent:
\begin{equation} \label{cumulative distribution2}
P_{cum}(\tilde{k})=\sum_{k'=\tilde{k}}^{\infty}P(k')\sim
\sum_{k'=\tilde{k}}^{\infty}e^{-k'/\kappa}\sim
e^{-\tilde{k}/\kappa}.
\end{equation}
This makes exponential distributions particularly easy to detect
experimentally, by plotting the corresponding cumulative
distributions on semilogarithmic scales.

Using equation~(\ref{Ki}), we have
$P_{cum}(k)=\sum_{k'=k}^{\infty}P(k')\\= P\left (t'\leq\tau=
t-\frac{k-1}{m} \right)$. Hence
\begin{eqnarray}\label{cumulative distribution3}
P_{cum}(k)&=&\sum_{t'=0}^{\tau}\frac{n_v(t')}{N_{t}}\nonumber \\
&=&\frac{2\,(1+m)^{ t-\frac{k-1}{m}}}{2\,(1+m)^{t}}=(1+m)^{-
\frac{k-1}{m}},
\end{eqnarray}
which decays exponentially with $k$. Note that when $m >1$, the
possible degrees are not arbitrary, equation~(\ref{cumulative
distribution3}) holds only for those $k$ being equal to 1 modulo
$m$. Thus the DURTs are a family of exponential networks, which have
a similar form of degree distribution as its stochastic version---
the URT~\cite{ZhZhZhGu08}.

\subsection{Average path length}

Average path length means the minimum number of edges connecting a
pair of nodes, averaged over all node pairs. It is defined to be:
\begin{equation}\label{APL01}
  \bar{d}_t  = \frac{S_t}{N_t(N_t-1)/2}\,,
\end{equation}
where $S_t$ denotes the sum of the total distances between two nodes
over all pairs, that is
\begin{equation}\label{APL02}
  S_t = \sum_{i\neq j} d_{i,j}\,,
\end{equation}
where $d_{i,j}$ is the shortest distance between node $i$ and $j$.
Note that in equation~(\ref{APL02}), for a couple of nodes $i$ and
$j$ ($i\neq j$), we only count $d_{i,j}$ or $d_{j,i}$, not both.

Let $\Omega_{\rm new}^{t}$ and $\Omega_{\rm old}^{t}$ represent the
sets of nodes created at step $t$ or earlier, respectively. Then one
can write the sum over all shortest paths $S_{t}$ in network $U_t$
as
\begin{equation}\label{APL03}
  S_{t} = \sum_{i \in \Omega_{\rm new}^{t},\,j\in \Omega_{\rm old}^{t}} d_{i,j}+\sum_{i \in \Omega_{\rm new}^{t},\,j\in
      \Omega_{\rm new}^{t}} d_{i,j}+\sum_{i \in \Omega_{\rm old}^{t},\,j\in
      \Omega_{\rm old}^{t}} d_{i,j},
\end{equation}
where the third term is exactly $S_{t-1}$, i.e.,
\begin{equation}\label{APL04}
  \sum_{i \in \Omega_{\rm old}^{t},\,j\in
      \Omega_{\rm old}^{t}} d_{i,j}=S_{t-1}.
\end{equation}
By construction, we can obtain the following relations for the first
and second terms on the right-hand side of equation~(\ref{APL03}):
\begin{equation}\label{APL05}
  \sum_{i \in \Omega_{\rm new}^{t},\,j\in
      \Omega_{\rm old}^{t}} d_{i,j}=m\,(N_{t-1}^{2}+2\,S_{t-1}),
\end{equation}
\begin{equation}\label{APL06}
  \sum_{i \in \Omega_{\rm new}^{t},\,j\in
      \Omega_{\rm new}^{t}} d_{i,j}=m^{2}\,S_{t-1}+m\,N_{t-1}(m\,N_{t-1}-1).
\end{equation}
The term $m\,N_{t-1}(m\,N_{t-1}-1)$ in equation~(\ref{APL06}) pops
out from counting: each path connecting two new points comes from a
path connecting two old points by adding two edges, is by increasing
the length by 2. As there are $\frac{1}{2}m\,N_{t-1}(m\,N_{t-1}-1)$
pairs of new points, the total increase in length is
$m\,N_{t-1}(m\,N_{t-1}-1)$.

Substituting equations.~(\ref{APL04}),~(\ref{APL05})
and~(\ref{APL06}) into equation~(\ref{APL03}) and considering
$N_{t}=2\,(1+m)^{t}$, the total distance is obtained to be
\begin{eqnarray}\label{APL07}
 S_t &=& (1+m)^{2}\,S_{t-1}+m\,(1+m)N_{t-1}^{2}-m\,N_{t-1}\nonumber \\
   &=& (1+m)^{2\,t}S_0 + m\,(1+m)\,\sum_{i=0}^{t-1} (1+m)^{2\,(t-1-i)} N_i^{2} \nonumber \\&\quad&-m\,\sum_{i=0}^{t-1} (1+m)^{2\,(t-1-i)} N_i\nonumber \\
&=& (4\,m\,t+m-1)\,(1+m)^{2\,t-1}+2\,(1+m)^{t-1}.
\end{eqnarray}
Inserting equation~(\ref{APL07}) into equation~(\ref{APL01}), we
have
\begin{eqnarray}\label{APL08}
  \bar{d}_t &=& \frac{2\,[(4\,m\,t+m-1)(1+m)^{2\,t-1}+2\,(1+m)^{t-1}]}{2\,(1+m)^{t}[2\,(1+m)^{t}-1]}\nonumber \\
  &=&\frac{(1+m)^{t}(4\,m\,t+m-1)+2}{2\,(1+m)^{t+1}-(1+m)}.
\end{eqnarray}
In the infinite network size limit ($t \rightarrow \infty$),
\begin{eqnarray}\label{APL09}
\bar{d}_{t} &\cong& \frac{2m}{m+1}t+\frac{m-1}{2(m+1)} \nonumber
\\&=& \frac{\ln N_{t}}{\ln (m+1)}-\frac{\ln 2}{\ln
(m+1)}+\frac{m-1}{2(m+1)},
\end{eqnarray}
which means that the average path length shows a logarithmic scaling
with the size of the network, indicating a similar small-world
behavior as the URT~\cite{DoMeOl06} and the Watts-Strogatz (WS)
model~\cite{WaSt98}.

\subsection{Betweenness distribution}

Betweenness of a node is the accumulated fraction of the total
number of shortest paths going through the given node over all node
pairs~\cite{Newman01,Ba04}. More precisely, the betweenness of a
node $i$ is
\begin{equation}
b_{i}=\sum_{j \ne i \neq k}\frac{\sigma_{jk}(i)}{\sigma_{jk}},
\end{equation}
where $\sigma_{jk}$ is the total number of shortest path between
node $j$ and $k$, and $\sigma_{jk}(i)$ is the number of shortest
path running through node $i$.

Since for a tree, there is a unique shortest path between each pair
of
nodes~\cite{SzMiKe02,BoRi04,GhOhGoKaKi04,ZhZhChGuFaZh07,ZhZhChGu08}.
Thus the betweenness of a node is simply given by the number of
distinct shortest paths passing through the node. Then in $U_t$, the
betweenness of a $\tau$-generation-old node $v$, which is created at
step $t-\tau+1$, denoted as $b_{t}(\tau)$ becomes
\begin{equation}\label{between01}
b_{t}(\tau)= \Theta_{t}^{\tau}\,\left[N_{t} -
\left(\Theta_{t}^{\tau}+1\right)\right]+\frac{\Theta_{t}^{\tau}(\Theta_{t}^{\tau}-1)}{2}
-\sum_{h=2}^{\tau-1} m\frac{\Theta_{t}^{h}(\Theta_{t}^{h}+1)}{2},
\end{equation}
where $\Theta_{t}^{\tau}$ denotes the total number of descendants of
node $v$ at time $t$, where the descendants of a node are its
children, its children's children, and so on. Note that the
descendants of node $v$ exclude $v$ itself. The first term in
equation~(\ref{between01}) counts shortest paths from descendants of
$v$ to other vertices. The second term accounts for the shortest
paths between descendants of $v$. The third term describes the
shortest paths between descendants of $v$ that do not pass through
$v$.

To find $b_{t}(\tau)$, it is necessary to explicitly determine the
descendants $\Theta_{t}^{\tau}$ of node $v$, which is related to
that of $v's$ children via~\cite{GhOhGoKaKi04,ZhZhChGuFaZh07}
\begin{equation}\label{child01}
\Theta_{t}^{\tau}=
\sum_{j=1}^{\tau-1}m\left(\Theta_{t}^{j}+1\right).
\end{equation}
Using $\Theta_{t}^{1}=0$, we can solve equation~(\ref{child01})
inductively,
\begin{equation}\label{child02}
\Theta_{t}^{\tau}= (1+m)^{\tau-1}-1.
\end{equation}
Substituting the result of equation~(\ref{child02}) and
$N_t=2\,(1+m)^{t}$ into equation~(\ref{between01}), we have
\begin{eqnarray}\label{between02}
b_{t}(\tau)&=&2\,(1+m)^{t+\tau-1} -2\,(1+m)^{t}\nonumber\\
&\quad& -\frac{(m+3)\,[(1+m)^{2\,(\tau-1)}-1]}{2\,(m+2)},
\end{eqnarray}
which is approximately equal to $2\,(m+1)^{t+\tau-1}$ for large
$\tau$. Then the cumulative betweenness distribution is
\begin{eqnarray}\label{pcumb01}
P_{\rm cum}(b)&=&\sum_{\mu \leq t-\tau+1}\frac{n_v(\mu)}{N_t}\nonumber\\
&=&\frac{(1+m)^{t+1}}{(1+m)^{t+\tau}} \approx {N_{t} \over b}\sim
b^{-1},
\end{eqnarray}
which shows that the betweenness distribution exhibits a power law
behavior with exponent $\gamma_{b}=2$, the same scaling has been
also obtained for the URT~\cite{GoOhJeKaKi02} and the $m=1$ case of
the Barab\'asi-Albert (BA) model~\cite{BaAl99} describing a random
scale-free treelike network~\cite{SzMiKe02,BoRi04}. Therefore,
power-law betweenness distribution is not an exclusive property of
scale-free networks.

\subsection{Degree correlations}

An interesting quantity related to degree correlations~\cite{MsSn02}
is the average degree of the nearest neighbors for nodes with degree
$k$, denoted as $k_{\rm nn}(k)$ \cite{PaVaVe01,VapaVe02,ZhZh07}.
When $k_{\rm nn}(k)$ increases with $k$, it means that nodes have a
tendency to connect to nodes with a similar or larger degree. In
this case the network is defined as assortative \cite{Newman02}. In
contrast, if $k_{\rm nn}(k)$ is decreasing with $k$, which implies
that nodes of large degree are likely to have near neighbors with
small degree, then the network is said to be disassortative. If
correlations are absent, $k_{\rm nn}(k)={\rm const}$.

For $U_t$, we can exactly calculate $k_{\rm nn}(k)$. Except for the
initial two nodes generated at step 0, no nodes born at the same
step, which have the same degree, will be linked to each other. All
links to nodes with larger degree are made at the creation step, and
then links to nodes with smaller degree are made at each subsequent
steps. This results in the expression for $k=1+m\,(t-t_i)$ ($t_i\geq
1$)
\begin{eqnarray}\label{knn01}
k_{\rm nn}(k)&=&{1\over n_v(t_i) k(t_i,t)} \Bigg[
  \sum_{t'_i=0}^{t'_i=t_i-1} m\,n_v(t'_i)k(t'_i,t)\nonumber\\
  &\quad&+\sum_{t'_i=t_i+1}^{t'_i=t} m\,n_v(t_i)
  k(t'_i,t)\Bigg],
\end{eqnarray}
where $k(t_i,t)$ represents the degree of a node at step $t$, which
was generated at step $t_i$. Here the first sum on the right-hand
side accounts for the links made to nodes with larger degree (i.e.\
$t'_i<t_i$) when the node was generated at $t_i$. The second sum
describes the links made to the current smallest degree nodes at
each step $t'_i>t_i$.

After some algebraic manipulations, equation~(\ref{knn01}) is
simplified to
\begin{eqnarray} \label{knn02}
k_{\rm nn}(k)&=&
\frac{2\,m\,t+2-2\,m\,t_i+m}{1+m\,(t-t_i)}+\frac{m^{2}\,(t-t_i-1)(t-t_i)}{2\,[1+m\,(t-t_i)]}
\nonumber\\ &\quad&-\frac{(1+m)^{1-t_i}}{1+m\,(t-t_i)}.
\end{eqnarray}
Writing equation (\ref{knn02}) in terms of $k$, it is
straightforward to obtain
\begin{eqnarray} \label{knn3}
k_{\rm nn}(k)=
\frac{k}{2}+\frac{2-m}{2}+\frac{3m+1}{2k}-\frac{(1+m)^{1+\frac{k-1}{m}}}{k\cdot
(1+m)^t}.
\end{eqnarray}
Thus we have obtained the degree correlations for those nodes born
at $t_i\geq 1$. For the initial two nodes, each has a degree of
$k=1+m\,t$, and it is easy to obtain
\begin{eqnarray}\label{knn4}
k_{\rm nn}(k=1+m\,t) &=&{1\over k}\, \left (\sum_{t'_i=1}^{t'_i=t}
  m\,k(t'_i,t) +k(0,t)\right)\nonumber\\
  &=&\frac{k}{2}+\frac{2-m}{2}+\frac{m-1}{2k}.
\end{eqnarray}
From equations~(\ref{knn3}) and (\ref{knn4}), it is obvious that for
large network (i.e., $t\rightarrow \infty$), $k_{\rm nn}(k)$ is
approximately a linear function of $k$, which shows that the network
is assortative.

\section{Eigenvalues and eigenvectors of the adjacency matrix}

As known from section 2, there are $2(m+1)^t$ vertices in $U_t$. we
denote by $V_t$ the vertex set of $U_t$, i.e.,
$V_t=\{v_1,v_2,\ldots, v_{2(m+1)^t}\}$. Let $\mathbf{A}_t=[a_{ij}]$
be the adjacency matrix of network $U_t$, where $a_{ij} =a_{ji}=1$
if nodes $i$ and $j$ are connected, $a_{ij} =a_{ji}=0$ otherwise.
For an arbitrary graph, it is generally difficult to determine all
eigenvalues and the corresponding eigenvectors of its adjacency
matrix, but below we will show that for $U_t$ one can settle this
problem.

\subsection{eigenvalues}

We begin by studying the eigenvalues of $U_t$. By construction, it
is easy to find that the adjacency matrix $\textbf{A}_t$ satisfies
the following relation:

\begin{equation}\label{matrix01}
\mathbf{A}_t=\left(\begin{array}{ccccc}\textbf{A}_{t-1} &
\textbf{I}& \textbf{I} &\cdots& \textbf{I}
\\\textbf{I} &{\textbf{0}}& {\textbf{0}}&\cdots& {\textbf{0}}
\\\textbf{I} & {\textbf{0}}& {\textbf{0}}&\cdots& {\textbf{0}}
\\\vdots & \vdots & \vdots & \  &\vdots
\\\textbf{I} & {\textbf{0}}& {\textbf{0}}&\cdots& {\textbf{0}}
\end{array}\right)_{(m+1)\times(m+1)}
\end{equation}
where each block is a $2(m+1)^{t-1}\times 2(m+1)^{t-1}$ matrix and
$\textbf{I}$ is identity matrix. Then, the characteristic polynomial
of $\textbf{A}_t$ is
\begin{eqnarray}\label{matrix04}
P_t(x)&=&{\rm det}(x\textbf{I}-\textbf{A}_t)\nonumber\\
&=&{\rm det}\left(\begin{array}{ccccc}x\textbf{I}-\textbf{A}_{t-1} &
-\textbf{I}& -\textbf{I} &\cdots& -\textbf{I}
\\-\textbf{I} & x\textbf{I} & {\textbf{0}}&\cdots& {\textbf{0}}
\\-\textbf{I} & {\textbf{0}}& x\textbf{I} &\cdots& {\textbf{0}}
\\\vdots & \vdots & \vdots & \  &\vdots
\\-\textbf{I} & {\textbf{0}}& {\textbf{0}}&\cdots& x\textbf{I}
\end{array}\right)\nonumber
\\&=&({\rm det}(x\textbf{I}))^m\cdot{\rm det}\left(\begin{array}{ccccc}(x-\frac{m}{x})\textbf{I}-\textbf{A}_{t-1}
& \textbf{0}& \textbf{0} &\cdots& \textbf{0}
\\-\frac{1}{x}\textbf{I} & \textbf{I} & {\textbf{0}}&\cdots& {\textbf{0}}
\\-\frac{1}{x}\textbf{I} & {\textbf{0}}& \textbf{I} &\cdots& {\textbf{0}}
\\\vdots & \vdots & \vdots & \  &\vdots
\\-\frac{1}{x}\textbf{I} & {\textbf{0}}& {\textbf{0}}&\cdots& \textbf{I}
\end{array}\right)\nonumber,\nonumber\\
\end{eqnarray}
where the elementary operations of matrix have been used. According
to the results in~\cite{Si00}, we have
\begin{eqnarray}\label{matrix05}
P_t(x)&=&\big({\rm det}(x\textbf{I})\big)^m \cdot {\rm
det}\Big(\big(x-\frac{m}{x}\big)\textbf{I}-\textbf{A}_{t-1}\Big).
\end{eqnarray}
Thus, $P_t(x)$ can be written recursively as follows:
\begin{equation}\label{matrix06}
P_t(x)=x^{2m(m+1)^{t-1}}\cdot P_{t-1}(\varphi(x)),
\end{equation}
where $\varphi(x)=x-\frac{m}{x}$. This recursive relation given by
equation~(\ref{matrix06}) is very important, from which we will
determine the complete eigenvalues of $U_t$ and their corresponding
eigenvectors. Notice that $P_{t-1}(x)$ is a monic polynomial of
degree $2(m+1)^{t-1}$, then the exponent of $\frac{m}{x}$ in
$P_{t-1}(\varphi(x))$ is $2(m+1)^{t-1}$, and hence the exponent of
factor $x$ in $P_t(x)$ is
\begin{equation}
2m(m+1)^{t-1}-2(m+1)^{t-1}=2(m-1)(m+1)^{t-1}.
\end{equation}
Consequently, $0$ is an eigenvalue of $\textbf{A}_t$, and its
multiplicity is $2(m-1)(m+1)^{t-1}$.

Notice that $U_t$ has $2(m+1)^t$ eigenvalues. We represent these
$2(m+1)^t$ eigenvalues as $\lambda^t_1,
\lambda^t_2,\dots,\lambda^t_{2(m+1)^t}$, respectively. For
convenience, we presume $\lambda^t_1 \le \lambda^t_2 \le \dots \le
\lambda^t_{2(m+1)^t}$, and denote by $AE_t$ the set of these
eigenvalues, i.e. $AE_t=\{\lambda^t_1,
\lambda^t_2,\dots,\lambda^t_{2(m+1)^t}\}$. All the eigenvalues in
set $AE_t$ can be divided into two parts. According to the above
analysis, $\lambda=0$ is an eigenvalue with multiplicity
$2(m-1)(m+1)^{t-1}$, which provide parts of the eigenvalues of
$A_t$. We denote by $AE^{'}_t$ the set of eigenvalues 0 of $U_t$,
i.e.
\begin{equation}
AE^{'}_t=\{\underbrace{0,0,0,\dots,0,0}_{2(m-1)(m+1)^{t-1}\mbox{}}\}
\end{equation}
It should be noted that here we neglect the distinctness of elements
in the set. The remaining $4(m+1)^{t-1}$ adjacency eigenvalues of
$U_t$ are determined by the equation $P_{t-1}(\varphi(x))=0$. Let
these $4(m+1)^{t-1}$ eigenvalues be $\tilde{\lambda}^t_1,
\tilde{\lambda}^t_2,\dots,\tilde{\lambda}^t_{4(m+1)^{t-1}}$,
respectively. For convenience, we presume $\tilde{\lambda}^t_1 \le
\tilde{\lambda}^t_2 \le \dots \le \tilde{\lambda}^t_{4(m+1)^{t-1}}$,
and denote by $AE^{*}_t$ the set of these eigenvalues, i.e.
$AE^{*}_t=\{\tilde{\lambda}^t_1,
\tilde{\lambda}^t_2,\dots,\tilde{\lambda}^t_{4(m+1)^{t-1}}\}$.
Therefore, the eigenvalue set of $U_t$ can be expressed as
$AE_t=AE^{'}_t \cup AE^{*}_t$.

From equation~(\ref{matrix06}), we have that for an arbitrary
element in $AE_{t-1}$, say $\lambda_{i}^{t-1} \in AE_{t-1}$, both
solutions of $x-\frac{m}{x}=\lambda_{i}^{t-1}$ are in $AE^{*}_t$. In
fact, equation $x-\frac{m}{x}=\lambda_{i}^{t-1}$ is equivalent to
\begin{equation}\label{matrix07}
x^2-\lambda_{i}^{t-1} x-m=0.
\end{equation}
We use notations $\tilde{\lambda}_{i}^{t}$ and
$\tilde{\lambda}_{i+2(m+1)^{t-1}}^{t}$ to represent the two
solutions of equation~(\ref{matrix07}), since they provide a natural
increasing order of the eigenvalues of $U_t$, which can be seen from
below argument. Solving this quadratic equation, its roots are
obtained to be $\tilde{\lambda}_{i}^{t}=r_1(\lambda_{i}^{t-1})$ and
$\tilde{\lambda}_{i+2(m+1)^{t-1}}^{t}=r_2(\lambda_{i}^{t-1})$, where
the function $r_1(\lambda)$ and $r_2(\lambda)$ satisfy

\begin{eqnarray}
r_1(\lambda)=\frac{1}{2}\left(\lambda-\sqrt{\lambda^2+4m}\right),\label{matrix08}\\
r_2(\lambda)=\frac{1}{2}\left(\lambda+\sqrt{\lambda^2+4m}\right).\label{matrix00}
\end{eqnarray}

Substituting each adjacency eigenvalue of $U_{t-1}$ into equations
(\ref{matrix08}) and~(\ref{matrix00}), we can obtain the set
$AE^{*}_{t}$ of eigenvalues of $U_{t}$. Since $AE_{0}=\{-1,1\}$, by
recursively applying the functions provided by equations
(\ref{matrix08}) and~(\ref{matrix00}), the eigenvalues of $U_{t}$
can be determined completely.

It is obvious that both $r_1(\lambda)$ and $r_2(\lambda)$ are
monotonously increasing functions. On the other hand, since
$r_1(\lambda)=\frac{1}{2}\left(\lambda-\sqrt{\lambda^2+4m}\right)=\frac{-2m}{\left(\lambda+\sqrt{\lambda^2+4m}\right)}$,
so $r_1(\lambda)<0$. Similarly, we can show that $r_2(\lambda)>0$.
Thus for arbitrary fixed $\lambda'$, $r_1(\lambda) < r_2(\lambda')$
holds for all $\lambda$. Then we have the following conclusion: If
the eigenvalues set of $U_{t-1}$ is
$AE_{t-1}=\{\lambda^{t-1}_1,\lambda^{t-1}_2,\dots,\lambda^{t-1}_{2(m+1)^{t-1}}\}$,
then solving equations~(\ref{matrix08}) and (\ref{matrix00}) we can
obtain the eigenvalue set $AE^{*}_t$ of $U_t$ to be
$AE^{*}_t=\{\tilde{\lambda}^t_1,
\tilde{\lambda}^t_2,\dots,\tilde{\lambda}^t_{4(m+1)^{t-1}}\}$, \
where $\tilde{\lambda}^t_1 \le \tilde{\lambda}^t_2 \le \dots \le
\tilde{\lambda}^t_{2(m+1)^{t-1}} < 0 <
\tilde{\lambda}^t_{2(m+1)^{t-1}+1} \le
\tilde{\lambda}^t_{2(m+1)^{t-1}+2} \le \dots \le
\tilde{\lambda}^t_{4(m+1)^{t-1}}$. Recall that $AE^{'}_t$ is consist
of $2(m-1)(m+1)^{t-1}$ elements, all of which are $0$, so we can
easily get the eigenvalue set of $U_t$ to be $AE_t=AE^{*}_t \cup
AE^{'}_t$.

\begin{figure}
\begin{center}
\includegraphics[width=0.3\linewidth,trim=10 0 50 0]{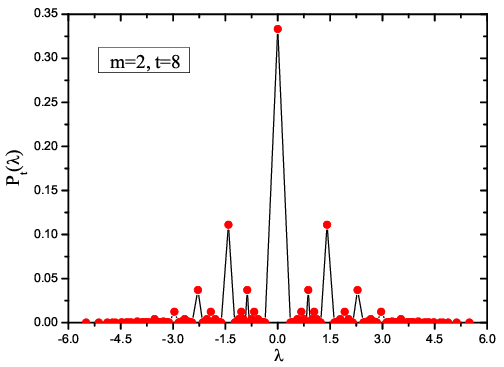}\\
\includegraphics[width=0.3\linewidth,trim=10 0 50 0]{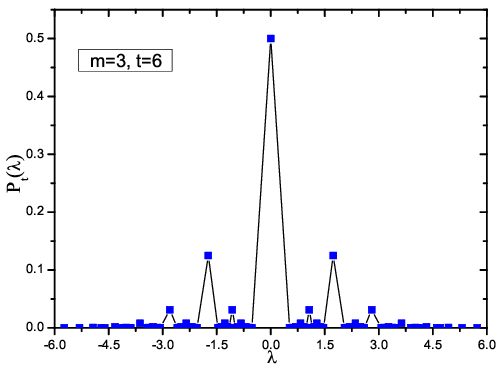}
\caption{The distribution of eigenvalues $P_t(\lambda)$ defined as
the ratio between the multiplicity of eigenvalue $\lambda$ and the
network order $N_t$.} \label{spectra}
\end{center}
\end{figure}

For above arguments, we can easily see that for the case of $m=1$,
all the $2^{t+1}$ eigenvalues of $U_t$ are different, which is an
interesting feature and has less been previously reported in other
network models thus may have some far-reaching consequences. For
other $m>1$, some eigenvalues multiple. In figure~\ref{spectra} we
plot the distribution of eigenvalues for two cases: $m=2$ and $m=3$.
It is observed that different from the uniform distribution of $m=1$
case, for $m>1$, the distribution of eigenvalues exhibit the form of
peaks.

%%%%%%%%%%%%%%%%%%%%%%%%%%%%%%%%%%%%%%%%%%%%%%%%%%%%%%%%%%
% Figure  2
%%%%%%%%%%%%%%%%%%%%%%%%%%%%%%%%%%%%%%%%%%%%%%%%%%%%%%%%%%
%%%%%%%%%%%%%%%%%%%%%%%%%%%%%%%%%%%%%%%%%%%%%%%%%%%%%%%%%%

\subsection{eigenvectors}

Similarly to the eigenvalues, the eigenvectors of $\textbf{A}_t$
follow directly from those of $\textbf{A}_{t-1}$. Assume that
$\lambda$ is an arbitrary eigenvalue of $U_t$, whose corresponding
eigenvector is $\textbf{\emph{v}} \in \textbf{R}^{2(m+1)^t}$, where
$\textbf{R}^{2(m+1)^t}$ represents the $2(m+1)^t$-dimensional vector
space. Then we can solve equation ($\lambda\, \textbf{I}
-\textbf{A}_t)\textbf{\emph{v}}=0$ to find the eigenvector
$\textbf{\emph{v}}$. We distinguish two cases: $\lambda \in
AE^{*}_t$ and $\lambda \in AE^{'}_t$, which will be addressed in
detail as follows:

For the first case $\lambda \in AE^{*}_t$, we can rewrite the
equation ($\lambda\, \textbf{I} -\textbf{A}_t)\textbf{\emph{v}}=0$
as
\begin{equation}\label{vector01}
\left(\begin{array}{ccccc}\lambda\textbf{I}-\textbf{A}_{t-1} &
-\textbf{I}& -\textbf{I} &\cdots& -\textbf{I}
\\-\textbf{I} &\lambda\textbf{I}& {\textbf{0}}&\cdots& {\textbf{0}}
\\-\textbf{I} & {\textbf{0}}& \lambda\textbf{I}&\cdots& {\textbf{0}}
\\\vdots & \vdots & \vdots & \  &\vdots
\\-\textbf{I} & {\textbf{0}}& {\textbf{0}}&\cdots& \lambda\textbf{I}
\end{array}\right)\left(\begin{array}{c}\textbf{\emph{v}}_1\\\textbf{\emph{v}}_2\\\textbf{\emph{v}}_3 \\\vdots\\ \textbf{\emph{v}}_{m+1}
\end{array}\right)=0,
\end{equation}
where vector $\textbf{\emph{v}}_i$ ($1\le i \le m+1$) are components
of $\textbf{\emph{v}}$. Equation~(\ref{vector01}) leads to the
following equations:
\begin{eqnarray}
\big(\lambda\textbf{I}_{t-1}-\textbf{A}_{t-1}\big)\textbf{\emph{v}}_1-\textbf{\emph{v}}_2-\dots-\textbf{\emph{v}}_{m+1}=\textbf{0},\label{vector02}
\\ -\textbf{\emph{v}}_1+\lambda \textbf{\emph{v}}_i=\textbf{0}\ \ \  (2\le i \le m+1).\label{vector03}
\end{eqnarray}
Resolve equation~(\ref{vector03}) to find
\begin{eqnarray}\label{vector04}
\textbf{\emph{v}}_i=\frac{1}\lambda\textbf{\emph{v}}_1 \ \  (2\le i
\le m+1).
\end{eqnarray}
Substituting equation~(\ref{vector04}) into
equation~(\ref{vector02}) we have
\begin{eqnarray}\label{vector05}
\left[\left(\lambda-\frac{m}\lambda\right)\textbf{I}-\textbf{A}_{t-1}\right]\textbf{\emph{v}}_1=0,
\end{eqnarray}
which indicates that $\textbf{\emph{v}}_1$ is the solution of
equation~(\ref{vector02}) while $\textbf{\emph{v}}_i$ ($2\le i\le
m+1$) are uniquely decided by $\textbf{\emph{v}}_1$ via
equation~(\ref{vector04}).

From equation~(\ref{matrix06}) in preceding subsection, it is clear
that if $\lambda$ is an eigenvalue of adjacency matrix
$\textbf{A}_t$, then $f(\lambda)=\lambda-\frac{m}\lambda$ must be
one eigenvalue of $\textbf{A}_{t-1}$. \big(Recall that if
$\lambda=\tilde{\lambda}_i^t \in AE^{*}_t$, then
$\varphi(\tilde{\lambda}_i^t)=\lambda_{i}^{t-1}$ for $i\le
2(m+1)^{t-1}$, or
$\varphi(\tilde{\lambda}_i^t)=\lambda_{i-2(m+1)^{t-1}}$ for
$i>2(m+1)^{t-1}$\big).  Thus, equation~(\ref{vector05}) together
with equation~(\ref{matrix06}) shows that $\textbf{\emph{v}}_1$ is
an eigenvector of matrix $\textbf{A}_{t-1}$ corresponding to the
eigenvalue $\lambda-\frac{m}{\lambda}$ determined by $\lambda$,
while
\begin{equation}
\textbf{\emph{v}}=\left(\begin{array}{ccc}\textbf{\emph{v}}_1
\\\textbf{\emph{v}}_2\\\textbf{\emph{v}}_3\\\vdots\\\textbf{\emph{v}}_{m+1} \end{array}\right)=
\left(\begin{array}{ccc}\textbf{\emph{v}}_1\\
\frac{1}{\lambda}\textbf{\emph{v}}_1\\
\frac{1}{\lambda}\textbf{\emph{v}}_1 \\ \vdots \\
\frac{1}{\lambda}\textbf{\emph{v}}_1\end{array}\right)
\end{equation}
is an eigenvector of $\textbf{A}_t$ corresponding to the eigenvalue
$\lambda$.

Since for the initial graph $U_0$, its adjacency matrix
$\textbf{A}_0$ has two eigenvalues -1 and 1 with respective
eigenvectors $(1,-1)^\top$ and $(1,1)^\top$.  By recursively
applying the above process, we can obtain all the eigenvectors
corresponding to $\lambda \in AE^{*}_t$.

For the second case of $\lambda \in AE^{'}_t$, where all
$\lambda=0$, the equation ($\lambda\, \textbf{I}
-\textbf{A}_t)\textbf{\emph{v}}=0$ can be recast as
\begin{equation}\label{vector11}
\left(\begin{array}{ccccc}-\textbf{A}_{t-1} & -\textbf{I}&
-\textbf{I} &\cdots& -\textbf{I}
\\-\textbf{I} &\textbf{0}& {\textbf{0}}&\cdots& {\textbf{0}}
\\-\textbf{I} & {\textbf{0}}& \textbf{0}&\cdots& {\textbf{0}}
\\\vdots & \vdots & \vdots & \  &\vdots
\\-\textbf{I} & {\textbf{0}}& {\textbf{0}}&\cdots& \textbf{0}
\end{array}\right)\left(\begin{array}{c}\textbf{\emph{v}}_1\\\textbf{\emph{v}}_2\\\textbf{\emph{v}}_3 \\\vdots\\ \textbf{\emph{v}}_{m+1}
\end{array}\right)=0,
\end{equation}
where vector $\textbf{\emph{v}}_i$ ($1\le i \le m+1$) are components
of $\textbf{\emph{v}}$. Equation~(\ref{vector11}) leads to the
following equations:
\begin{eqnarray}
\textbf{\emph{v}}_1=\textbf{0},\label{vector12}
\\\textbf{\emph{v}}_2+\textbf{\emph{v}}_3+\dots+\textbf{\emph{v}}_{m+1}=\textbf{0}.\label{vector13}
\end{eqnarray}
From equation~(\ref{vector12}), $\textbf{\emph{v}}_1$ is a zero
vector, and we denote by  $\textbf{\emph{v}}_{i,j}$ the $j$-th
component of column vector $\textbf{\emph{v}}_i$.
Equation~(\ref{vector13}) gives us the following equations:
\[ \left \{
\begin{array}{ccccccccc}
\textbf{\emph{v}}_{2,1}&+&\textbf{\emph{v}}_{3,1}&+&\dots
&+&\textbf{\emph{v}}_{m+1,1}=\textbf{0}
\\\textbf{\emph{v}}_{2,2}&+&\textbf{\emph{v}}_{3,2}&+&\dots
&+&\textbf{\emph{v}}_{m+1,2}=\textbf{0}
\\ \vdots & \vdots  &\vdots  & \vdots  &\  &\vdots &\vdots
\\\textbf{\emph{v}}_{2,2(m+1)^{t-1}}&+&\textbf{\emph{v}}_{3,2(m+1)^{t-1}}&+&\dots
&+&\textbf{\emph{v}}_{m+1,2(m+1)^{t-1}}=\textbf{0}
\end{array} \right.
\]

The set of all solutions to any equation above consists of vectors
that can be written as:
\begin{equation}\label{vector15}
\left(\begin{array}{ccc}\textbf{\emph{v}}_{2,j}
\\\textbf{\emph{v}}_{3,j}\\\textbf{\emph{v}}_{4,j}\\\vdots\\\textbf{\emph{v}}_{m+1,j} \end{array}\right)=
k_{1,j}
\left(\begin{array}{ccc}\textbf{-1}\\
\textbf{1}\\
\textbf{0} \\ \vdots \\
\textbf{0} \end{array}\right)+ k_{2,j}
\left(\begin{array}{ccc}\textbf{-1}\\
\textbf{0}\\
\textbf{1} \\ \vdots \\
\textbf{0} \end{array}\right)+\dots+k_{m-1,j}
\left(\begin{array}{ccc}\textbf{-1}\\
\textbf{0}\\
\textbf{0} \\ \vdots \\
\textbf{1} \end{array}\right),
\end{equation}
where $k_{1,j}$ , $k_{2,j}$ , $\dots$ , $k_{m-1,j}$ are any real
numbers. From equation~(\ref{vector15}), the solutions for all the
vectors $\textbf{\emph{v}}_i$ ($2\le i \le m+1$) can be rewritten
as:
\begin{equation}\label{vector16}
\left(\begin{array}{c}\textbf{\emph{v}}_2^\top\\\textbf{\emph{v}}_3^\top\\\textbf{\emph{v}}_4^\top
\\\vdots\\ \textbf{\emph{v}}_{m+1}^\top
\end{array}\right)=
\left(\begin{array}{cccc}\textbf{-1} & \textbf{-1}& \cdots&
\textbf{-1}
\\\textbf{1} & {\textbf{0}}&\cdots& {\textbf{0}}
\\\textbf{0} & {\textbf{1}}& \cdots& {\textbf{0}}
\\\vdots & \vdots &  \  &\vdots
\\\textbf{0} & {\textbf{0}}& \cdots& \textbf{1}
\end{array}\right)\\
\\
\left(\begin{array}{cccc}k_{1,1} & k_{1,2}& \cdots&
k_{1,2(m+1)^{t-1}}
\\k_{2,1} &k_{2,2} & \cdots & k_{2,2(m+1)^{t-1}}
\\k_{3,1} & k_{3,2}& \cdots& k_{3,2(m+1)^{t-1}}
\\\vdots  & \vdots & \  &\vdots
\\k_{m-1,1}  & k_{m-1,2} &\cdots& k_{m-1,2(m+1)^{t-1}}
\end{array}\right),
\end{equation}
where $k_{i,j}$ \big($1\le i\le m-1$; $1\le j\le 2(m+1)^{t-1}$\big)
are arbitrary real numbers. According to the
equation~(\ref{vector16}), we can obtain the eigenvector
$\textbf{\emph{v}}$ corresponding to the eigenvalue $\textbf{0}$.
Moreover, it is easy to see that the dimension of the eigenspace of
matrix $A_t$ associated with eigenvalue $\textbf{0}$ is
$2(m-1)(m+1)^{t-1}$.

\section{Conclusion and discussion}

In conclusion, we have studied a family of deterministic models for
the uniform recursive tree, which we name the deterministic uniform
recursive trees (DURTs) that are constructed in a recursive way. The
DURTs are in fact deterministic variants of the intensively studied
random uniform recursive tree. We have presented an exhaustive
analysis of various structural properties of the DURTs, and obtained
the precise solutions for these features that include degree
distributions, average path length, betweenness distribution, and
degree correlations. Aside from their deterministic structures, the
obtained structural characteristics of the DURTs are similar to
those of URT. Consequently, the DURTs may provide useful insight to
the practices as URT.

Furthermore, by using the methods of linear algebra and graph
theory, we have performed a detailed analysis of the complete
eigenvalues and their corresponding eigenvectors of the adjacency
matrix for DURTs. We have fully characterized the spectral
properties and eigenvectors for DURTs. We have shown that all the
eigenvalues and eigenvectors of the adjacency matrix for DURTs can
be directly determined from those for the initial network. It is
expected that the methods applied here can be extended to a larger
type of deterministic networks.

\section*{Acknowledgment}

We would like to thank Shuyang Gao for preparing this manuscript.
This research was supported by the National Basic Research Program
of China under grant No. 2007CB310806, the National Natural Science
Foundation of China under Grant Nos. 60704044, 60873040 and
60873070, Shanghai Leading Academic Discipline Project No. B114, and
the Program for New Century Excellent Talents in University of China
(NCET-06-0376).

%%%%%%%%%%%%%%%%%%%%%%%%%%%%%%%%%%%%%%%%%%%%%%%%%%%%%%%%%%%%%%%%%
%%%%%%%%%%%%%%%%%%%%%%%%%%%%%%%%%%%%%%%%%%%%%%%%%%%%%%%%%%%%%%%%%

\end{document}